\documentclass[useAMS,twocolumn,usenatbib,usegraphicx]{mn2e}
\usepackage{graphicx}
\newcommand{\go}{\mathrel{\raise.3ex\hbox{$>$}\mkern-14mu
             \lower0.6ex\hbox{$\sim$}}}
\newcommand{\lo}{\mathrel{\raise.3ex\hbox{$<$}\mkern-14mu
             \lower0.6ex\hbox{$\sim$}}}

\newcommand{\vp}{v^{\prime}}
\newcommand{\vr}{v_{\rm r}}
\newcommand{\gammar}{\gamma_{\rm r}}

\newcommand{\vecr}{\bmath r}
\newcommand{\vecB}{\bmath B}
\newcommand{\vecm}{\bmath m}
\newcommand{\vecE}{\bmath E}
\newcommand{\vecn}{\bmath n}
\newcommand{\vecW}{\bmath \Omega}
\newcommand{\vecv}{\bmath v}
\newcommand{\veca}{\bmath a}
\newcommand{\vecvp}{\vecv^{\prime}}
\newcommand{\vecvr}{\vecv_{\rm r}}
\newcommand{\vecb}{\bmath b}

\newcommand{\vechatr}{\hat{\bmath r}}
\newcommand{\vechatm}{\hat{\bmath m}}
\newcommand{\vechatb}{\hat{\bmath b}}
\newcommand{\vechatv}{\hat{\bmath v}}

\newcommand{\vechata}{\hat{\bmath a}}
\newcommand{\vechate}{\hat{\bmath \epsilon}}
\newcommand{\vechatZ}{\hat{\bmath Z}}

\newcommand{\PA}{\phi_{\rm PA}}
\newcommand{\PAi}{\phi_{\rm PA\,0}}
\newcommand{\intd}{\rm d}

\title{Curvature Radiation in Rotating Pulsar Magnetosphere}

\author[P. F. Wang, C. Wang, and J. L. Han]
{P. F. Wang\thanks{E-mail: pfwang@nao.cas.cn}, C. Wang, and J. L. Han \\
  National Astronomical Observatories, Chinese Academy of
  Sciences.  A20 Datun Road, Chaoyang District, Beijing 100012, China \\
}

\pagerange{\pageref{firstpage}--\pageref{lastpage}}

\begin{document}

\maketitle

\label{firstpage}

\begin{abstract}
We consider the curvature emission properties from relativistic
particles streaming along magnetic field lines and co-rotating with
pulsar magnetosphere. The co-rotation affects the trajectories of the
particles and hence the emission properties, especially the
polarization. We consider the modification of the particle velocity
and acceleration due to the co-rotation. Curvature radiation from a
single particle is calculated using the approximation of a circular
path to the particle trajectory. Curvature radiation from particles at
a given height actually contains the contributions from particles
streaming along all the nearby field lines around the tangential
point, forming the emission cone of $1/\gamma$. The polarization
patterns from the emission cone are distorted by the additional
rotation, more serious for emission from a larger height. Net circular
polarization can be generated by the density gradient in the emission
cone. For three typical density models in the form of core, cone and
patches, we calculate the polarization profiles for emission generated
at a given height. We find that the circular polarization could have a
single sign or sign reversal, depending on the density gradient along
the rotation phase. The polarization profiles of the total curvature
radiation from the whole open field line region, calculated by adding
the emission from all possible heights, are similar to that from a
dominating emission height. The circular polarization of curvature
radiation has sign reversals in the patchy emission, while it has a
single sign for the core emission, and is negligible for the cone
emission.
\end{abstract}


\section{INTRODUCTION}

Pulsar radio emission is generally believed to be generated by
relativistic particles streaming out along the open magnetic field
lines in pulsar magnetosphere. The observed radio emission are
generally highly linearly polarized, and have significant circular
polarization \citep{lm88,hmx+98,rr03}. Various radio emission
mechanisms are suggested to explain polarization profiles
\citep[e.g.][]{bgi88,xlh+00,gan10}.

Curvature radiation serves as one of the most possible emission
mechanisms for pulsar radio emission. \citet{bb76} firstly developed
the general formalism for coherent curvature radiation by a
relativistic plasma streaming along curved trajectories using antenna
mechanism, assuming the plasma is perturbed by a plane
wave. \citet{bb77} applied this mechanism for pulsar radio emission
and calculated the luminosity. \citet{ou80} analytically calculated
the pulsar emission spectrum from curvature radiation. In polarization
aspect, it is known that curvature radiation is almost completely
linearly polarized. \citet{gs90} pointed out the sign reversal feature
for circular polarization (CP) of core emission. \citet{gan10}
considered the geometry of the emission region in pulsar
magnetosphere, obtained the polarization states self-consistently, and
explained the correlation between the position angle (PA) and the sign
reversal of CP \citep{hmx+98}.

Because pulsar magnetosphere is co-rotating with the neutron star, the
streaming particles should have co-rotation velocity additional to the
streaming velocity along the magnetic field line, which affects the
curvature radiation properties. \citet{bcw91} first proposed a
relativistic model for pulsar polarization by incorporating the
rotation effect, and predicted the phase lag between the centers of
the PA curve and the profile. The rotation also influences the
intensities of the leading and trailing components of the profile
\citep{bcw91,tg07}. \citet{dwd10} analyzed the rotation and found two
competing effects: the enhancement of the trailing side caused by
aberration and retardation (AR) and the weakening of the trailing side
caused by the rotation-induced asymmetry of curvature radius. In the
pioneer paper, \citet{bcw91} calculated only the emission of particles
with velocity pointing towards the observer, and did not consider the
emissions from particles of nearby field lines, which may still beam
towards us.  Meanwhile, the influence of rotation on polarization
intensities are not taken into account. \citet{gan10} considered the
emission from all nearby field lines, but did not incorporate the
rotation in his detailed work.

In this paper, we study the curvature radiation in the rotating pulsar
magnetosphere. The curvature emission of single particle and a bunch
of particles streaming along the magnetic field lines are calculated
using numerical simulations. The paper is organized as following. In
Section 2, we present the details of our calculation for the emission
from a single relativistic particle. In Section 3 the rotation
distorted patterns of curvature emission are calculated for the
emission cone at a given height. The polarization profiles from a
given height are calculated for three typical density models in the
form of core, cone and patches for various emission geometries. The
polarized profiles from the whole magnetosphere are presented in
Section 4. Our conclusions are given Section 5.

%%%%%%%%%%%%%%%%%%%%%%%%%%%%%%%%%%%%%%%%%%%%%%%%%%%%%%%%%%%%%%%%%%%%%%%%%%%%%%
%%%%%%%%%%%%%%%%%%%%%%%%%%%%%%%%%%%%%%%%%%%%%%%%%%%%%%%%%%%%%%%%%%%%%%%%%%%%%%
%%%%%%%%%%%%%%%%%%%%%%%%%%%%%%%%%%%%%%%%%%%%%%%%%%%%%%%%%%%%%%%%%%%%%%%%%%%%%%

\section{Curvature radiation of a single particle in rotating pulsar magnetosphere}

The pulsar magnetosphere consists of relativistic plasma streaming
along magnetic field lines. The magnetic field is generally
thought to be an inclined dipole, which is given by
\begin{equation}
\vecB=B_{\star}(\frac{R_{\star}}{r})^3[3\vechatr(\vechatr\cdot
\vechatm)-\vechatm],
\label{eq:staticb}
\end{equation}
where $B_{\star}$ is the surface magnetic field, $R_{\star}$ is the
neutron star radius, $\vechatr$ is the unit vector along $\vecr$, and
$\vechatm$ represents the unit vector for the magnetic dipole
moment. Curvature radiation is generated when relativistic charged
particles (electrons or positrons) streaming along the curved field
lines, the electric field from a single particle is
\begin{equation}
\vecE(t)=\frac{\vecn\times[(\vecn-\vecv)\times
\veca]}{(1-\vecn\cdot \vecv)^3},
\label{eq:E_t}
\end{equation}
with $\vecn$ the wave vector unity, $\vecv$ the particle velocity and
$\veca$ the acceleration. In previous studies of curvature radiation
in pulsar magnetospheres \citep{bb77, gan10}, the rotation of
magnetosphere is usually neglected, thus the particle velocity $\vecv$
is simply parallel to the tangential direction of the $\mathbf{B}$
field at the emission point, and the acceleration $\veca$ is in the
same $\mathbf{B}$ field line plane as the emission point. However, the
co-rotation of a particle with the magnetosphere changes $\vecv$ and
$\veca$, and affects the intensity and polarization of the particle
emission significantly \citep{bcw91}. In this section, we discuss the
correction to the particle velocity and acceleration in detail, and
give a precise equation to calculate the curvature radiation of a
single particle in rotating pulsar magnetosphere.

%%%%%%%%%%%%%%%%%%%%%%%%%%%%%%%%%%%%%%%%%%%%%%%%%%%%%%%%%%%%%%%%%%%%%%%%%%%%%%
%%%%%%%%%%%%%%%%%%%%%%%%%%%%%%%%%%%%%%%%%%%%%%%%%%%%%%%%%%%%%%%%%%%%%%%%%%%%%%
\subsection{The particle velocity, acceleration and emission location}

\begin{figure}
    \centering
    \includegraphics[angle=0, width=0.47\textwidth] {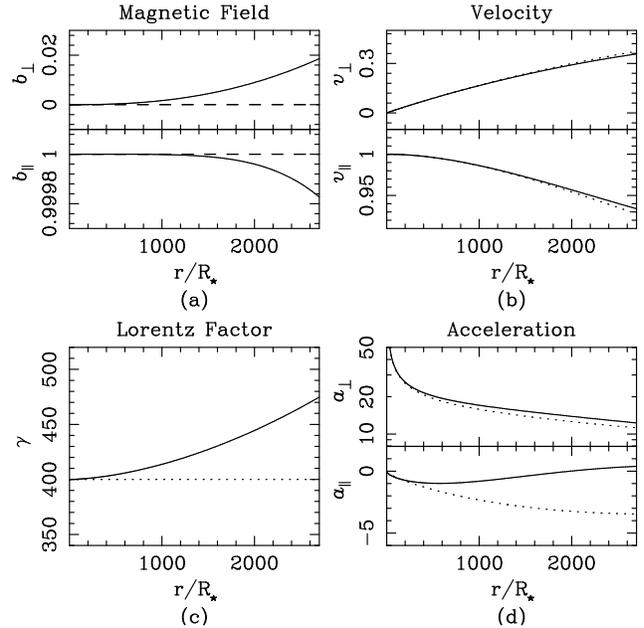}
    \caption{{\bf The modified magnetic field direction, velocity,
        Lorentz factor and acceleration of the particles along the
        last open field line considering the co-rotation of pulsar
        magnetosphere. Fig. (a) shows the magnetic field directions in
        the fixed lab frme (solid line) and the co-rotating frame
        (dashed line). In Fig. (b), (c) and (d), the solid lines show
        the precise relativistic velocity, Lorentz factor and
        acceleration given by equation~(\ref{eq:vel}) and (\ref{eq:acc}),
        the dotted lines are approximations given by
        equations~(\ref{eq:cv}) and (\ref{eq:a_appro}). Our calculation
        chooses the last open field line, which is in a plane with
        azimuthal angle of $-10^\circ$. The other parameters involved
        in calculation are $P=1s$, $\alpha=70^{\circ}$, and
        $\gamma=400$.}}
    \label{fig:va}
\end{figure}

In the co-rotating frame, the relativistic particles or bunches are
streaming along the open magnetic field lines with a velocity of
\begin{equation}
\vecvp=\vp \hat{\bmath b^\prime}
\label{eq:vp}
\end{equation}
here $\hat{\bmath b^\prime}=\bmath B/|\bmath B|$ is a unit vector
along the tangent direction of the magnetic field line, $\vp$ is the
velocity magnitude in terms of light speed $c$. The particles also
experience rotation with velocity $\vecvr=\vecW\times\vecr/c$. In the
lab frame, the total velocity $\vecv$ should be the addition of
$\vecvp$ with $\vecvr$ by Lorentz transformation, which reads
\begin{equation}
\vecv=\frac{\vecvp+\frac{(\gammar-1)}{\vr^2} (\vecvr \cdot \vecvp)
\vecvr + \gammar \vecvr} {\gammar (1+ \vecvr \cdot \vecvp)}
\label{eq:vel}
\end{equation}
here $\vr$ is the magnitude for the rotation velocity. Its
corresponding Lorentz factor is $\gammar=1/\sqrt{1-\vr^2}$.
Equation~(\ref{eq:vel}) gives the precise particle velocity in the lab
frame. {\bf The velocity $\vecv$ can also be written as
\begin{equation}
\vecv = \kappa \vechatb + \vecW\times\vecr/c.
\label{eq:cv}
\end{equation}
Here, $\kappa$ is the function of $\vecv$ and $\vecvr$, $\vechatb$
denotes the magnetic field direction in the fixed lab frame, which is
the Lorentz transformation of $\hat{\bmath b^\prime}$. The difference
between $\hat{\bmath b}$(solid line) and $\hat{\bmath
  b^\prime}$(dashed line) is very small for a low emission height
($r<500 R_\star$) as shown in Fig.~\ref{fig:va} (a). Therefore, most
of the previous studies \citep{bcw91,kg12} can ignore the difference
and use $\hat{\bmath b^\prime}$ instead of $\hat{\bmath b}$ in
equation~(\ref{eq:cv}). In \citet{bcw91}, $\kappa$ is limited to be a
constant, since they just consider the emission from low heights where
$\vr$ is very small. To coordinate the total velocity with the light
speed, \citet{gan05} introduced the evolvement of $\kappa$ with height
by assuming the Lorentz factor to be a constant, as shown with the
dotted line Fig.~\ref{fig:va}(c).  However, according to
equation~(\ref{eq:vel}), the Lorentz factor for the particle is changing
with height shown by the solid line in Fig.~\ref{fig:va}(c). The
precise relativistic velocity given by equation~(\ref{eq:vel}) and the
approximation are shown in Fig.~\ref{fig:va}(b). Within several
hundred of $R_\star$, the difference is very small. }

\begin{figure}
    \centering
    \includegraphics[angle=0, width=0.45\textwidth] {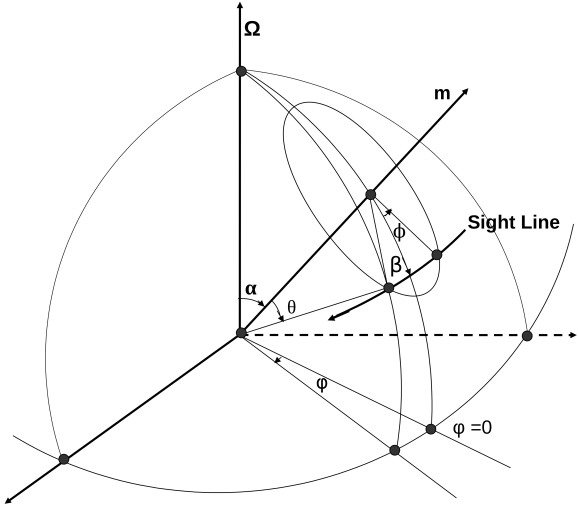}
    \caption{The coordinates for pulsar emission geometry. $\bmath
      \Omega$ represents the rotation axis, $\bmath m$ is the dipole
      moment. The magnetic inclination angle is $\alpha$ with respect
      to $\bmath \Omega$. Sight line has impact angle of $\beta$,
      cutting the beam. The rotation phase is $\varphi=\Omega t$ starting
      from the $\bmath \Omega-\bmath m$ plane. The magnetic polar and
      azimuthal angles are $\theta$ and $\phi$.}
\label{fig:geometry}
\end{figure}
\begin{figure}
    \centering
    \includegraphics[angle=0, width=0.25\textwidth] {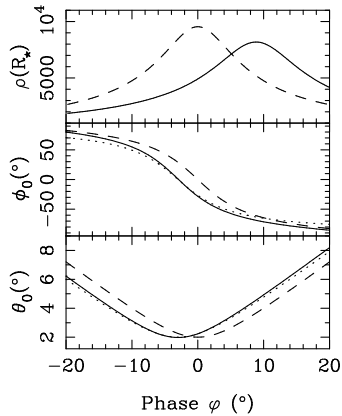}
    \caption{Tangential emission points (magnetic polar angle
      $\theta_0$ and azimuth angle $\phi_0$) and the corresponding
      curvature radii $\rho$ for various rotation phases, $\varphi$. The
      dashed lines represent the case without rotation, while the
      solid lines are the results by taking the rotation into
      account. The dotted lines show the results given by
      \citet{bcw91}. The parameters used for calculation are $P=1s$,
      $\alpha=30^{\circ}$, $\beta=3^{\circ}$, $\gamma=400$, $r=250
      R_{\star}$.}
\label{fig:location}
\end{figure}

The radiation of a relativistic particle is beamed in its velocity
direction $\vechatv$. It requires the alignment between the wave
vector unity $\vecn$ and $\vechatv$ for the observer to receive
considerable radiation. At a given emission height, the emission
location ($\theta_0$, $\phi_0$) at each rotation phase $\varphi$ (see
Fig.~\ref{fig:geometry}) can be determined by solving
$\vecn=\vechatv$, which is the tangential point of the trajectory
along the line of sight. Here ($\theta_0$, $\phi_0$) is the polar
coordinate of magnetic axis, with $\phi_0=0$ corresponding to the
$\vecW$-$\vecm$ plane. \citet{bcw91} has done similar calculations by
analytically solving $\vecn=\kappa \vechatb + \vecW\times\vecr/c$. Our
numerical calculations develop their results further to higher
magnetosphere using equation~(\ref{eq:vel}). The emission locations from
our calculation are shown in Fig.~\ref{fig:location}, which is very
close to that given by \citet{bcw91}. Rotation demonstrates its effect
by shifting the emission locations towards an early phase.

The particle acceleration is also affected by the co-rotation with
pulsar magnetosphere. The charged particle streams along the magnetic
field line, whose trajectory is described by $r=r_e\sin^2\theta$ in
the co-rotating magnetic axis frame. Here $r_e$ is the field line
constant, $\theta$ is the polar angle from the magnetic axis. The
particle acceleration in the co-rotating frame reads
\begin{equation}
\veca^{\prime}= \frac{v^{\prime
2}c}{|\vecb|}\frac{\partial\vechatb}{\partial\theta}
\end{equation}
here $\vecb=\partial\vecr/\partial\theta$ is the magnetic field line
tangent and its magnitude
$|\vecb|=(r_e/\sqrt{2})\sin\theta\sqrt{5+3\cos(2\theta)}$. The
acceleration in the lab frame can be calculated by
\begin{equation}
\veca=\frac{\intd \vecv}{\intd t}.
\label{eq:acc}
\end{equation}
Substitute the approximation of $\vecv$ (equation~{\ref{eq:cv}) in the
  above equation, we have
\begin{equation}
\veca=\frac{\kappa^2 c}{|\vecb|}\frac{\partial{\vechatb}}
{\partial{\theta}}+\frac{\kappa c}{|\vecb|}\frac{\partial{\kappa}}
{\partial{\theta}}\vechatb+ 2 \kappa \vecW\times \vechatb +
\vecW\times(\vecW\times\vecr/c),
\label{eq:a_appro}
\end{equation}
{\bf which is equivalent to equation~(7) in \citet{kg12}.}  Substitute the
precise velocity equation~(\ref{eq:vel}) in equation~(\ref{eq:acc}), we can also
obtain the analytic equation of $\veca$ without approximation, which
is too complicated to write done here. The numerical results for the
precise acceleration from Lorentz transformation and the approximated
one given by equation~(\ref{eq:a_appro}) are shown in
Fig.~\ref{fig:va}(d). The approximated treatment for the acceleration
remains good only at low emission height of several hundred
$R_{\star}$, but becomes significantly different from the precise
acceleration at outer magnetosphere.

The curvature radius of the particle trajectory in the lab frame is
approximately given by $\rho \simeq c v^2/a$. In the upper panel of
Fig.~\ref{fig:location}, we calculated the curvature radii for the
trajectories at each rotation phase using the acceleration. The
leading side tends to have smaller curvature radius compared to the
case without rotation, while the trailing side behaves in the opposite
way. \citet{tg07} pointed out that the rotation induces a significant
curvature into the particle trajectories, which are deflected towards
the rotation direction. The magnitude of the deflection on leading
side becomes larger compared with the no-rotating case, while the
deflection on the trailing side is weakened. Our results is consistent
with \citet{tg07}. The modifications of particle velocity,
acceleration and emission location due to the co-rotation of particles
with pulsar magnetosphere have significant roles on pulsar radiation
property, which will be discussed in Section 3.

%%%%%%%%%%%%%%%%%%%%%%%%%%%%%%%%%%%%%%%%%%%%%%%%%%%%%%%%%%%%%%%%%%%%%%%%%%%%%%
%%%%%%%%%%%%%%%%%%%%%%%%%%%%%%%%%%%%%%%%%%%%%%%%%%%%%%%%%%%%%%%%%%%%%%%%%%%%%%
\subsection{Curvature Emission from a single particle}

\begin{figure}
    \centering
    \includegraphics[angle=0, width=0.4\textwidth] {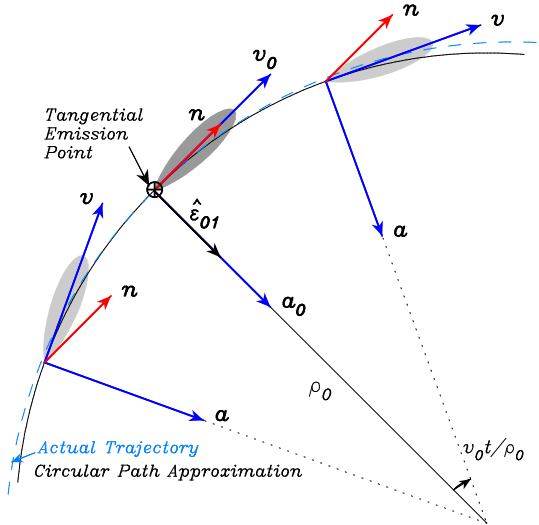}
    \caption{The circular path approximation (solid curve) to the
      actual particle trajectory (dashed curve).  At the tangential
      emission point, the particle velocity $\vecv_0$ is parallel to
      the line of sight $\vecn$. }
\label{fig:cirpathsketch}
\end{figure}
Knowing the particle velocity, acceleration and emission location of
the relativistic particle, we can directly calculate its curvature
radiation fields. Note that according to equation~(\ref{eq:E_t}), the
emission location, where $\vecn\cdot\vechatv=1$, is the point where
the radiation electric field is the strongest. Before and after this
point within the angle of $1/\gamma$, the particle still emits in the
direction of $\vecn$ but the magnitude of the electric field becomes
weakened (see Fig.~\ref{fig:cirpathsketch}). The total electric field
emitted from this particle should be the coherent integration of the
emission from the whole trajectory, especially near the tangential
emission point.

\begin{figure}
    \centering
    \includegraphics[angle=0, width=0.3\textwidth] {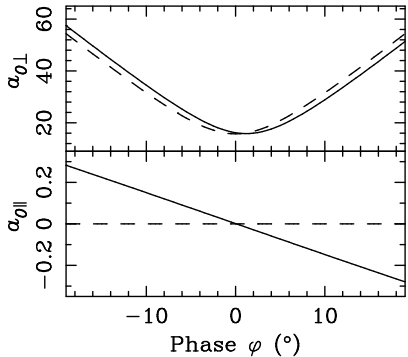}
    \caption{The components of acceleration at each emission point
      parallel and perpendicular to the velocity direction. The solid
      lines are for the corrected acceleration including the rotation
      effect, while the dashed lines are drawn without considering the
      rotation. The parameters used here are $P=1s$,
      $\alpha=30^{\circ}$, $\beta=3^{\circ}$, $\gamma=400$, $r=50
      R_{\star}$.}
\label{fig:a}
\end{figure}

To study the emission from a relativistic particle, we set up a
reference frame ($\vechate_{01}$, $\vechate_{02}$, $\vechatv_0$) at
the tangential point along the ling of sight in the instantaneous
trajectory plane, shown in Fig.~\ref{fig:cirpathsketch}.  Here the
subscript "$_0$" stands for the parameters at the tangential point,
$\vecv_0$ and $\veca_0$ are the velocity and acceleration at the
tangential point, $\vechatv_0$ is the unit vector along $\vecv_0$. The
other two axes of the frame are described as
\begin{eqnarray}
\vechate_{01} &=& \vechate_{02}\times\vechatv_0,   \nonumber\\
\vechate_{02} &=& \frac{\vechatv_0\times\vechata_0}{\mid
\vechatv_0\times\vechata_0 \mid},
\end{eqnarray}
$\vechate_{01}$ is in the $\vecv_0$-$\veca_0$ plane and almost
parallel to $\veca_0$, and $\vechate_{02}$ is perpendicular to the
plane. In Fig.~\ref{fig:a}, we compare the two orthogonal components
of $\veca_0$ along $\vechatv_0$ ($\veca_{0\parallel}$) and
$\vechate_{01}$ ($\veca_{0\perp}$), and find
$|\veca_{0\parallel}|/|\veca_{0\perp}| \lo 10^{-2}$ for various
particle trajectories at the corresponding pulsar rotation phases.
This means that the contribution of $\veca_{0\parallel}$ to the total
emission can be neglected. It was pointed out by \citet{jac75} that
the radiation emitted by an extremely relativistic particle subject to
arbitrary accelerations is equivalent to that by a particle moving on
an appropriate circular path. The approximate circular path around the
tangential point for the particle trajectory has a radius of curvature
of
\begin{equation}
\rho_0=c\frac{v^2}{|\veca_0|},
\end{equation}
here we suppose $\veca_0\simeq \veca_{0\perp}$, since
$|\veca_{0\parallel}| \ll |\veca_{0\perp}|$. The velocity and
acceleration of the bunch in the vicinity of the tangential point
can be expressed by
\begin{eqnarray}
\vecv &=& |\vecv_0| [\cos(\frac{v_0 t}{\rho_0}) \vechatv_0 +
 \sin(\frac{v_0 t}{\rho_0})\vechate_{01}],     \nonumber\\                                               \nonumber\\
\veca &=& |\veca_0| [-\sin(\frac{v_0 t}{\rho_0}) \vechatv_0 +
\cos(\frac{v_0 t}{\rho_0}) \vechate_{01} ].
\end{eqnarray}
The position vector for the particle in the circular path reads
\begin{equation}
\vecr_{\rm cir}=-\rho_0 \vechata,
\end{equation}
with $\vechata=\veca/|\veca|$. The radiating electric field of the
relativistic particle at frequency $\omega$ in direction $\vecn$
is the Fourier Transform of $\vecE(t)$, which is governed by
\begin{equation}
\vecE(\omega)=\frac{q e^{i \omega R_0/c}}{\sqrt{2 \pi} c R_0}
\int_{-\infty}^{\infty} \frac{\vecn\times[(\vecn-\vecv)\times
\veca]}{(1-\vecn\cdot \vecv)^2} e^{i\omega (t-\vecn\cdot
\vecr_{\rm cir}/c)} \intd t,
\label{eq:Ew}
\end{equation}
here, $R_0$ is the distance between the circular path center and the
observer.

We should note that the integration in equation~\ref{eq:Ew} is highly
oscillating and not easy to integrate. Following \citet{gan10}, we
expand the integrand to the third order around $t=0$ and then get the
analytic expression of the integration. The total integration can then
be calculated directly from the analytic equation. The detailed
description of this method is given in the Appendix B of
\citet{gan10}.

\begin{figure}
    \centering
    \includegraphics[angle=0,width=0.45\textwidth] {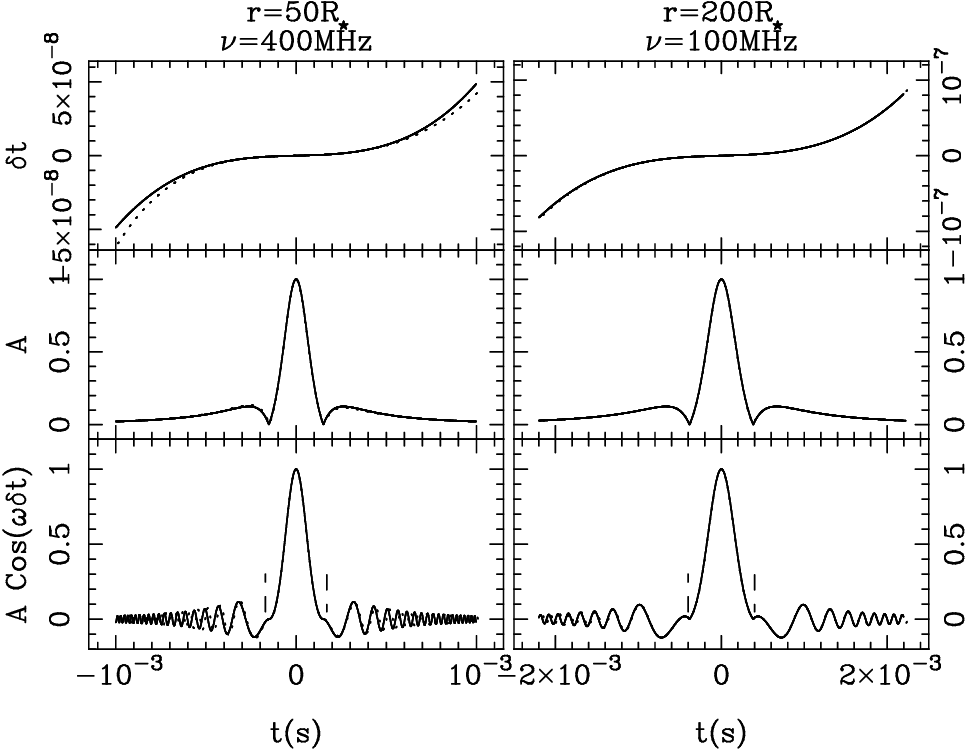}
    \caption{The integrand term in equation~(\ref{eq:Ew}) near the
      tangential emission point (where $t=0s$) considering the
      co-rotation of the particle with pulsar magnetosphere. Here
      $A=|\vecn\times[(\vecn-\vecv)\times\veca]/(1-\vecn\cdot\vecv)^2|$,
      $\delta t=t-\vecn\cdot \vecr_{\rm cir}/c$, and
      $A\cos(\omega\delta t)$ are the magnitude of the integrand. The
      dotted lines are for the actual trajectory of the particle in
      the meridional plane, while the solid lines present the results
      using the circular path approximation. The left and right panels
      show the results for lower and higher emission radius (and
      corresponding frequencies), respectively. The other parameters
      used for calculation here are $P=1s$, $\alpha=10^{\circ}$,
      $\beta=3^{\circ}$ and $\gamma=400$.}
\label{fig:appro-comp}
\end{figure}

In our calculation, we use the assumption of a circular path to
describe the trajectory of the relativistic particles around the
tangential emission point. To demonstrate its feasibility, we further
compare in Fig.~\ref{fig:appro-comp} the integrand term in
equation~(\ref{eq:Ew}) near the tangential emission point using actual
trajectory, {\bf similarly done by \cite{kg12}} and circular path
approximation. The magnitude of the integrand is highly oscillating
when the particle is away from the tangential emission point about
$|t|>4\times 10^{-4}s$ (or about $10R_\star$), so that the integration
outside is almost vanished due to the oscillation. The central part
near $t=0$ between the two dashed lines in the figure dominates the
result of integration. In this central part, the difference by using
the assumed circular path and the ideal trajectory is
negligible. Therefore, the electrical fields calculated from the
circular path approximation are accurate enough but much easier.

The observed emission is the transverse component of the electric
field perpendicular to the line of sight $\vecn$. We have to transform
the calculated $\vecE(\omega)$ to the lab frame $XYZ$, with $\vechatZ$
along the line of sight $\vecn$, and $\vecW$ in the $XZ$ plane. The
Stokes parameters are defined as follows
\begin{eqnarray}
I &=& E_XE_X^\ast+E_YE_Y^\ast,         \nonumber\\
Q &=& E_XE_X^\ast-E_YE_Y^\ast,         \nonumber\\
U &=& 2 {\rm{Re}}[E_X^\ast E_Y],                    \nonumber\\
V &=& 2 {\rm{Im}}[E_X^\ast E_Y].
\end{eqnarray}

%%%%%%%%%%%%%%%%%%%%%%%%%%%%%%%%%%%%%%%%%%%%%%%%%%%%%%%%%%%%%%%%%%%%%%%%%%
%%%%%%%%%%%%%%%%%%%%%%%%%%%%%%%%%%%%%%%%%%%%%%%%%%%%%%%%%%%%%%%%%%%%%%%%%%
\subsection{Pulse profile of curvature emission from a single particle}

\begin{figure}
    \centering
    \includegraphics[angle=0, width=0.45\textwidth] {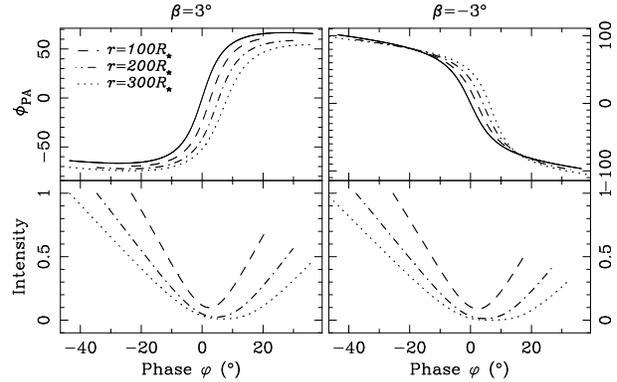}
    \caption{Pulse profiles of intensity and position angles emitted
      from a single particle at the tangential emission point at
      various rotation phase. The calculation is made for different
      heights $r=100 R_{\star}$, $200 R_{\star}$ and $300
      R_{\star}$. The solid line presents the PA profile without
      rotation. The other parameters used here are $P=1s$,
      $\alpha=30^{\circ}$ {\bf and $\gamma=400$}.}
\label{fig:papad}
\end{figure}

Previously, only the radiation exactly along the velocity direction of
the relativistic particle is considered for the pulsar profile. It
means that only the radiation by a particle at ($\theta_0$, $\phi_0$)
with a velocity towards the observer contributes to the observed
intensity at a given rotation phase. Using equation~(\ref{eq:Ew}), we can
calculate the polarized emission from a single particle at various
phases to get the intensity and polarization profiles, as is shown in
Fig.~\ref{fig:papad}. The results are very similar to those given by
\citet{bcw91}, where they use the total emission of the particle in
the emission cone at the tangential point instead of the integration
along the trajectory. The emission is almost 100\% linearly polarized
since the emission at each phase comes from a single particle. The
leading part of the intensity profile is brighter than the trailing
side because of the rotation induced curvature for the particle
trajectory. The most obvious are the polarization position angle
curves shifted towards a later phase due to the aberration effect.

\begin{figure}
    \centering
    \includegraphics[angle=0, width=0.45\textwidth] {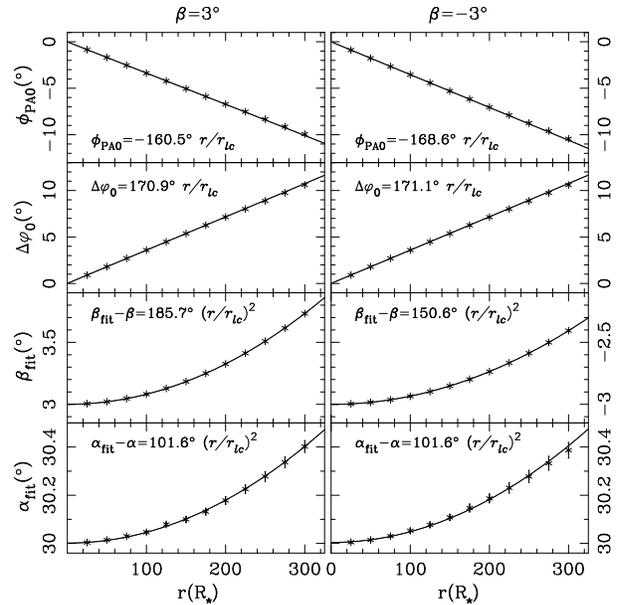}
    \caption{The fitted parameters $\alpha_{\rm fit}$, $\beta_{\rm
        fit}$, $\Delta\varphi_0$ and $\PAi$ vary with emission heights. The
      dots are the values by fitting the PA profiles shown in
      Fig.~\ref{fig:papad} with the rotating vector model
      (equation~\ref{eq:rvm}). The solid lines present the best fit to
      the dots, which is described by the equations given in each
      panel.}
\label{fig:pafit}
\end{figure}

The shift of a PA profile caused by rotation is larger at a larger
emission height due to the relatively larger rotation velocity
$\vecW\times\vecr$. To show the shifts varying with the emission
heights, we can fit the shifted PA curves at different heights by
using the rotating vector model \citep{kmc70}
\begin{eqnarray}
  \lefteqn{\tan(\PA-\PAi)=}                         \nonumber\\
  &&\frac{\sin (\varphi-\Delta\varphi_0) \sin
\alpha}{\sin(\alpha+\beta) \cos \alpha-\cos (\alpha+\beta) \sin
\alpha \cos (\varphi-\Delta\varphi_0)}.
\label{eq:rvm}
\end{eqnarray}

Here $\PAi$ describes the PA profile shift along the perpendicular
direction in Fig.~\ref{fig:papad}, $\Delta\varphi_0$ represents the phase
shift of a PA profile. Evolution of the four fitting parameters
($\alpha_{\rm fit}$, $\beta_{\rm fit}$, $\Delta\varphi_0$ and $\PAi$)
with height are shown in Fig.~\ref{fig:pafit}. Rotation induced
aberration not only delays the polarization sweep with positive
$\Delta\varphi_0$, but also shifts the entire PA sweep downward with
negative $\PAi$ \citep{ha01}. Both of these shifts ($\PAi$ and
$\Delta\varphi_0$) are the first order functions of $r$. Pulsar emission
geometry is also distorted by the rotation, since the fitting values
of $\alpha_{\rm fit}$ and $\beta_{\rm fit}$ change with heights. PA
curves are steepened for $\beta<0$ but flattened for $\beta>0$
\citep{bcw91}. The fitting value of the dipole inclination angle
$\alpha_{\rm fit}$ increases with $r$. The influences of rotation on
$\alpha_{\rm fit}$ and $\beta_{\rm fit}$ determination are both second
order functions of $r$, as $\beta_{\rm fit}-\beta\propto r^2$ and
$\alpha_{\rm fit}-\alpha\propto r^2$.

%%%%%%%%%%%%%%%%%%%%%%%%%%%%%%%%%%%%%%%%%%%%%%%%%%%%%%%%%%%%%%%%%%%%%%%%%%%%%%
%%%%%%%%%%%%%%%%%%%%%%%%%%%%%%%%%%%%%%%%%%%%%%%%%%%%%%%%%%%%%%%%%%%%%%%%%%%%%%
%%%%%%%%%%%%%%%%%%%%%%%%%%%%%%%%%%%%%%%%%%%%%%%%%%%%%%%%%%%%%%%%%%%%%%%%%%%%%%
\section{Curvature radiation at a given height}

\begin{figure*}
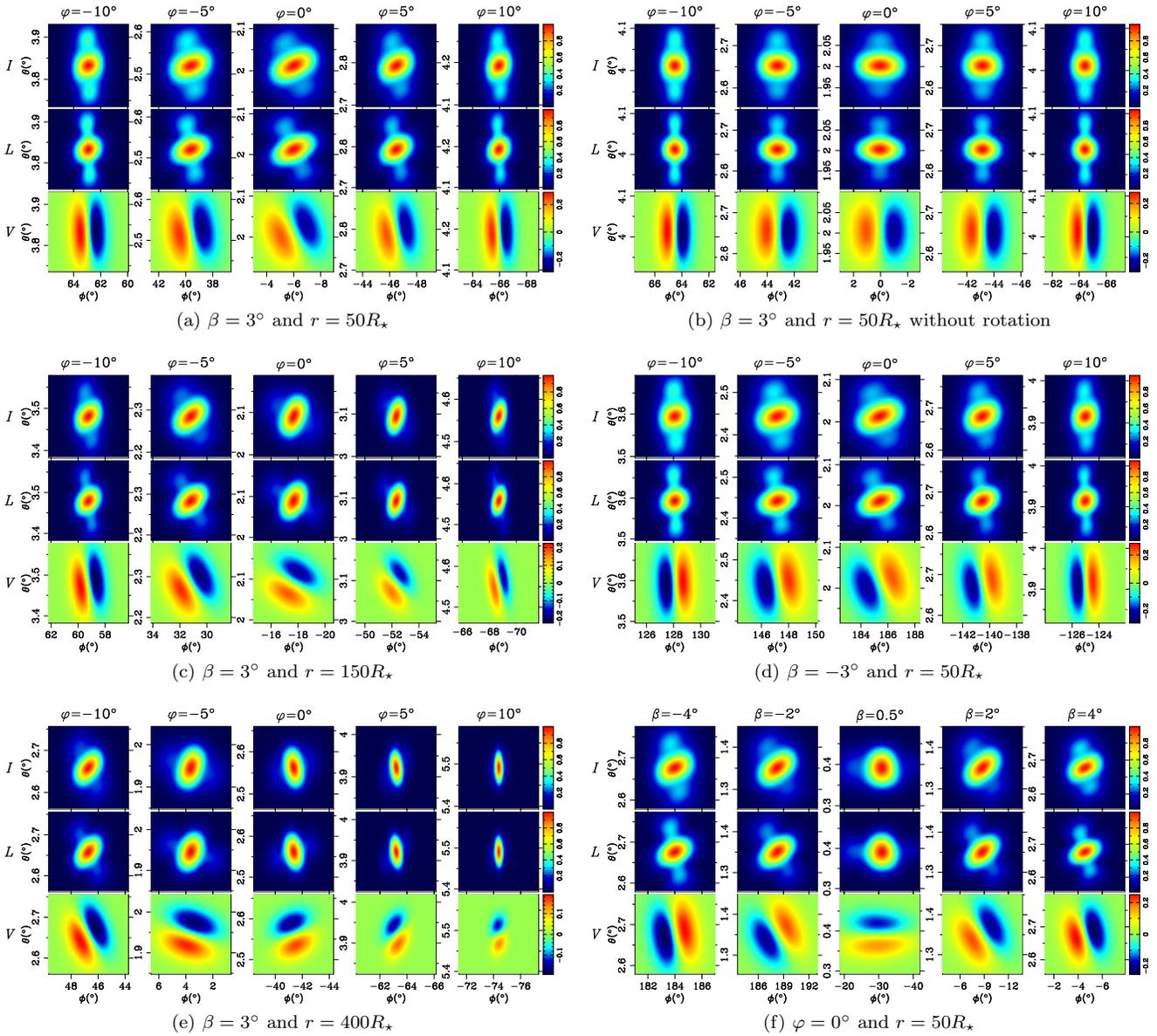

\begin{tabular}{cc}
\includegraphics[angle=0, width=0.48\textwidth] {minibeamr50.ps} &
\includegraphics[angle=0, width=0.48\textwidth] {minibeamrno.ps} \\
(a) $\beta=3^{\circ}$ and $r=50R_{\star}$
&
(b) $\beta=3^{\circ}$ and $r=50R_{\star}$ without rotation            \\
\\
\includegraphics[angle=0, width=0.48\textwidth] {minibeamr150.ps} &
\includegraphics[angle=0, width=0.48\textwidth] {minibeamr50_beta-.ps} \\
(c) $\beta=3^{\circ}$ and $r=150R_{\star}$ &
(d) $\beta=-3^{\circ}$ and $r=50R_{\star}$          \\
\\
\includegraphics[angle=0, width=0.48\textwidth] {minibeamr400.ps} &
\includegraphics[angle=0, width=0.48\textwidth] {minibeamr50betas-n.ps} \\
(e) $\beta=3^{\circ}$ and $r=400R_{\star}$  &
(f) $\varphi=0^\circ$ and $r=50R_{\star}$            \\
\end{tabular}
\caption{Emission patterns for the total intensity $I$, linear
  polarization $L$ and circular polarization $V$ around the tangential
  emission point calculated for various heights and $\beta$ at
  different rotation phases ($\varphi=0^\circ$, $\pm5^\circ$,
  $\pm10^\circ$). The patterns without rotation are always similar for
  different $r$ and $\beta$, so that we just show one for
  $r=50R_{\star}$ and $\beta=3^{\circ}$. The two axes for plots are
  the magnetic polar angle $\theta$ and azimuth angle $\phi$. The
  other parameters used for calculations are $P=1s$,
  $\alpha=30^{\circ}$, $\gamma=400$ and {\bf $\nu=\omega/2\pi=400$ MHz}.}
\label{fig:pattern}
\end{figure*}
A relativistic particle streaming along a magnetic field line in
magnetosphere will emit in a cone of $1/\gamma$ around its velocity
direction. Thus the observed emission at a pulsar rotation phase comes
from the particles not only along one magnetic field line with
$\vecn\cdot\vechatb=1$, but also the nearby field lines within the $1/
\gamma$ emission cone. In this section, we will calculate the
curvature radiation from the particles at a given height steaming
along a bunch of magnetic field lines.

%%%%%%%%%%%%%%%%%%%%%%%%%%%%%%%%%%%%%%%%%%%%%%%%%%%%%%%%%%%%%%%%%%%%%%%%%%%%%%
%%%%%%%%%%%%%%%%%%%%%%%%%%%%%%%%%%%%%%%%%%%%%%%%%%%%%%%%%%%%%%%%%%%%%%%%%%%%%%
\subsection{Emission pattern at a given height}

The emission received at one rotation phase actually contains the
contributions from all nearby field lines around the tangential
emission point at a given height. Fig.~\ref{fig:pattern} shows the
emission patterns for the polarized emission with or without rotation
at different heights and impact angle $\beta$. The patterns are
plotted in the coordinates ($\theta$, $\phi$) around the magnetic axis
for different rotation phases. The central point of the patterns
corresponds to the tangential point where the velocity aligns with the
line of sight. The patterns extend to approximately $0.15^\circ$ in
magnetic polar angle $\theta$, and about $2^\circ$--$4^\circ$ in
magnetic azimuth $\phi$, which agree with the $1/\gamma$ emission cone
since we take $\gamma=400$. The total intensity $I$ distributes in an
ellipse manner around the central maximum. The linear polarization $L$
has almost the same patterns as $I$ but with smaller magnitudes. The
circular polarization $V$ is shown in two antisymmetrical lobes with
$+$ and $-$ signs corresponding to the left and right hands, with the
maximum intensity of $V$ as about 20\% of the peak total intensity.

Our calculations in Fig.~\ref{fig:pattern} make it clear that the
rotation systematically distorts the emission patterns compared with
the case without rotation in which the long axes of $I$ and $L$ beams
and the zero line of the $V$ are all vertically aligned. Since
particles at a larger emission height have comparatively larger
rotation velocity and hence trajectories are much more bent. For the
emission from higher radii, the distortion of beam pattern due to
rotation becomes more serious. The beam patterns for opposite impact
angles (see Fig.~\ref{fig:pattern}a and \ref{fig:pattern}d for
$\pm\beta$) are almost the same except for the opposite signs in the
$V$ patterns. The beam pattern for a smaller impact angle $\beta$
tends to have large distortion (see Fig.~\ref{fig:pattern}f), because
the rotation induced acceleration is more dominating for the particles
in the field lines near the magnetic axis, compared to the curvature
acceleration. Note that the beam patterns for the phases with opposite
signs, i.e., $\varphi=\pm 5^\circ$ or $\pm10^\circ$, are
asymmetric. The strongest distortion happens at a late phase
($\varphi>0^\circ$), because the rotation bends the trajectories
towards the rotating direction. The long axes of the beam patterns for
an extremely high emission radii, i.e., $r \sim 400R_\star$ or higher,
are rotated by more than $90^\circ$ compared those without rotation.

%%%%%%%%%%%%%%%%%%%%%%%%%%%%%%%%%%%%%%%%%%%%%%%%%%%%%%%%%%%%%%%%%%%%%%%%%%%%%%
%%%%%%%%%%%%%%%%%%%%%%%%%%%%%%%%%%%%%%%%%%%%%%%%%%%%%%%%%%%%%%%%%%%%%%%%%%%%%%
\subsection{Polarization profiles for the emission from a given height}

The observed Stokes parameters at each rotation phase are the
integration of all visible beam patterns from all particles in all
field lines. It is generally believed that pulsar radiations are
coherent, and the width of the coherent bunches in the $\theta$-$\phi$
plane should be much smaller than the radiating wavelength for the
coherency to be efficient. The typical size of the $1/\gamma$ emission
pattern at a given height is much larger than the coherent
width. Therefore, the emission at a given height should be added
incoherently, which reads
\begin{eqnarray}
I_r &=& \int^{\theta_0+\delta\theta}_{\theta_0-\delta\theta}
\int^{\phi_0+\delta\phi}_{\phi_0-\delta\phi}
 N(r,\theta,\phi)~I~r^2 \sin \theta \intd \theta \intd \phi,         \nonumber\\
Q_r &=& \int^{\theta_0+\delta\theta}_{\theta_0-\delta\theta}
\int^{\phi_0+\delta\phi}_{\phi_0-\delta\phi}
 N(r,\theta,\phi)~Q~r^2 \sin \theta \intd \theta \intd \phi,         \nonumber\\
U_r &=& \int^{\theta_0+\delta\theta}_{\theta_0-\delta\theta}
\int^{\phi_0+\delta\phi}_{\phi_0-\delta\phi}
 N(r,\theta,\phi)~U~r^2 \sin \theta \intd \theta \intd \phi,         \nonumber\\
V_r &=& \int^{\theta_0+\delta\theta}_{\theta_0-\delta\theta}
\int^{\phi_0+\delta\phi}_{\phi_0-\delta\phi}
 N(r,\theta,\phi)~V~r^2 \sin \theta \intd \theta \intd \phi.
\label{eq:stokes_r}
\end{eqnarray}
Here $N(r,\theta,\phi)$ serves as particle density distribution
function, the half-width $\delta\theta$ and $\delta\phi$ of the
emission pattern are chosen to be $0.1^{\circ}$ and $2^{\circ}$ to
cover the pattern of the emission cone. For simplicity we assume that
the density distributions across $r$, $\theta$ and $\phi$ are
independent, i.e., $N(r,\theta,\phi)=n(r)f(\theta)g(\phi)$. The
density variation along the height $n(r)\propto r^{-3}$, and $n(r)$ is
a constant for a given emission height.

\begin{table}
\small %
\begin{center}
\caption{The sign of the net circular polarization of particle
  emission for various density gradient at an emission height of $50
  R_{\star}$.}
\label{tab:Vsign}
%\scriptsize
\begin{tabular}{c|cc}
\hline
                   &  $\beta>0$ & $\beta<0$ \\
\hline
\hline
 $\partial N/\partial \theta>0$ &  $-$  & $+$   \\
 $\partial N/\partial \theta<0$ &  $+$  & $-$   \\
\hline
 $\partial N/\partial \phi>0$   &  $+$  & $+$   \\
 $\partial N/\partial \phi<0$   &  $-$  & $-$   \\
\hline
\end{tabular}
\end{center}
\end{table}

If the particle density, $N$, keeps constant in all field lines at a
given height, the observed circular polarization should be zero in all
the cases since it is always symmetrical with the central point, and
the integrations of particle emission from many field lines will smear
out the circular polarization. However, in most cases the density
distribution across field lines are not uniform. Modulation of $N$
across $\theta$ and $\phi$ may result in net circular
polarization. For example, in the case without rotation in
Fig.~\ref{fig:pattern}b, the pattern of $V$ is symmetric with
respect to $\phi$, so that only the density gradient along $\phi$ can
produce net $V$, while the density gradient along $\theta$ can
not. The detailed discussion for this case can be found in
\citet{gan10}. In the cases of rotation, the observed beam patterns
become complicated due to the distortion. The circular polarization
depends on the density gradients along $\theta$ and $\phi$, following
the rule shown in Table~\ref{tab:Vsign} for the emission from a
typical emission height $r=50R_{\star}$ (see Fig.~\ref{fig:pattern}a
and f).

%%%%%%%%%%%%%%%%%%%%%%%%%%%%%%%%%%%%%%%%%%%%%%%%%%%%%%%%%%%%%%%%%%%%%%%%%%%%%%
\subsubsection{Cone-core model}

\begin{figure}
    \centering
    \includegraphics[angle=0, width=0.45\textwidth] {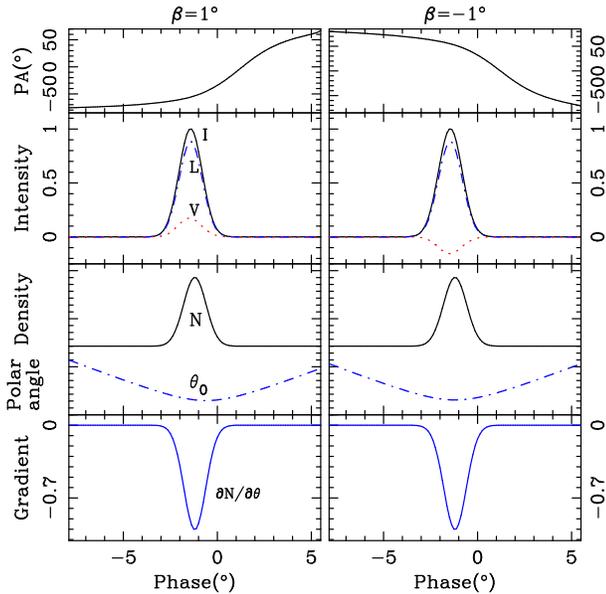}
      \caption{Polarization profiles ($I$, $L$, $V$ and PA) at 400 MHz
        from a height of $50 R_{\star}$ for the core density
        distribution. The solid, dash-dotted and dotted lines
        represent $I$, $L$ and $V$, respectively. The density $N$,
        magnetic polar angle, $\theta$, and the density gradient
        $\partial N /\partial \theta$ vary with the rotation phase
        $\varphi$. The left and right panels are plotted for opposite
        impact angles $\beta=\pm 1^\circ$. The other parameters used
        here are $P=1s$, $\alpha=30^{\circ}$, $\gamma=400$ and
        $\sigma_{\theta_\star}=0.0005$.}
\label{fig:core}
\end{figure}

The observation of pulsar mean profiles suggested a classical pulsar
beam model with a central core surrounded by one or multiple hollow
cones \citep{bac76,ran93}. These cone-core regions in pulsar
magnetosphere are filled with relativistic particles streaming along
the open magnetic field lines, which is generated by sparking process
on neutron star polar cap \citep{rs75}. We explore the emission
profile for the particle density distributions in the form of core and
cone in the polar cap region streaming out from the neutron star
surface and extending to high magnetosphere.

The core density distribution is written as
\begin{eqnarray}
f(\theta)&=&f_0 \exp \left(-\frac{\theta_\star^2}{2 \sigma_{\theta_\star}^2}\right),   \nonumber\\
g(\phi)&=&1.
\label{eq:core_dis}
\end{eqnarray}
Here $f_0$ is the maximum density, $\theta_\star$ is the polar angle
of a field line footed on the neutron star surface in the polar cap
region, $\sigma_{\theta_\star}$ is the characteristic width of the
gaussian distribution on the star surface. The polarized profiles from
particles distributed in this core at a height of $r=50R_{\star}$ are
shown in Fig.~\ref{fig:core}. Due to rotation, the ``S'' shaped PA
curve is shifted towards a later phase, while the intensity curve to
an earlier phase, as predicted by \citet{bcw91}. Single sign of
circular polarization is generated because of the density gradient
along $\theta$. In the case of positive $\beta$ (see the left panels
of Fig.~\ref{fig:core}), the left hand CP ($V>0$) is generated because
$\partial N/\partial \theta<0$ for all phases, but the right hand CP
for negative $\beta$ (see the right panels of
Fig.~\ref{fig:core}). Both of them agree well with
Table~\ref{tab:Vsign}. The core density model always has $\partial
N/\partial \theta<0$ and $\partial N/\partial \phi=0$, so that the net
CP always has a single sign for all rotation phases.

\begin{figure}
    \centering
    \includegraphics[angle=0, width=0.45\textwidth] {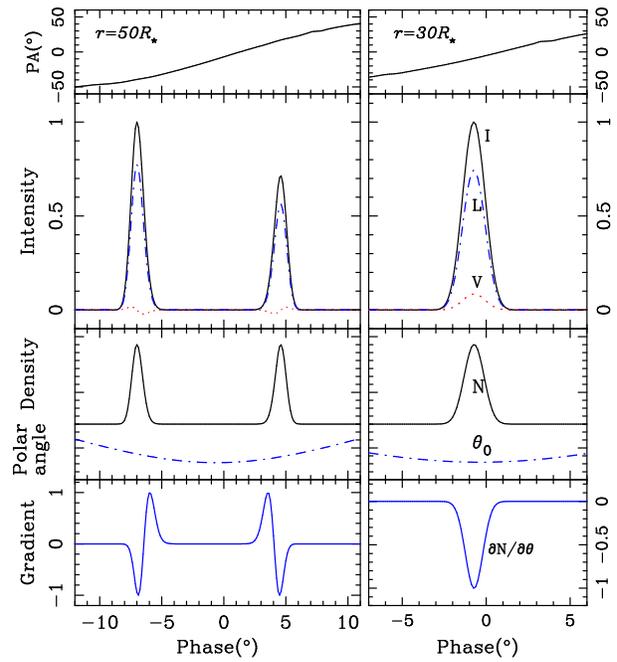}
    \caption{Same as Fig.~\ref{fig:core} but for the density
      distribution in the form of a cone. Left and right panels stand
      for the emission profiles from 30 $R_{\star}$ and 50
      $R_{\star}$, respectively. The other parameters used here are
      $P=1s$, $\alpha=30^{\circ}$, $\beta=5^{\circ}$, $\gamma=400$,
      $\sigma_{\theta_\star}=0.0002$ and $s=0.7$.}
\label{fig:cone}
\end{figure}

The cone density distribution reads
\begin{eqnarray}
f(\theta)&=&f_0 \exp \left[-\frac{(\theta_\star - s \theta_{\star
{\rm
lof}})^2} {2 \sigma_{\theta_\star}^2}\right],  \nonumber\\
g(\phi)&=&1.
\label{eq:cone_dis}
\end{eqnarray}
Here $\theta_{\star{\rm lof}} = \sin^{-1} (\sqrt{R_{\star}/r_{e,{\rm
      lof}}})$ is the polar angle maximum on the star surface
corresponding to the last open field line, the dimensionless parameter
$s$ describes the peak position for the maximum density in terms of
$\theta_{\star{\rm lof}}$. The two sets of polarized profiles for the
cone density model are shown in Fig.~\ref{fig:cone} with the same
impact angle of $\beta=5^\circ$ but two different emission
heights. For a higher emission radius of $r=50R_\star$ (left panels),
the line of sight cuts through the density cone close to the magnetic
axis, which results in double intensity components. For each
component, $\partial N/\partial \theta$ changes signs, which results
in the sign reversal of circular polarization. The opposite $\partial
N/\partial \theta$ variation causes the opposite sign reversals of
$V$. For a lower emission height of $r=30R_\star$ (right panels),
sight line will cut across the edge part of the density
cone. Therefore only one component can be observed. The density
gradient $\partial N/\partial \theta<0$, so that the circular
polarization has one sign across the pulse phases, very similar
as the case of core emission.

%%%%%%%%%%%%%%%%%%%%%%%%%%%%%%%%%%%%%%%%%%%%%%%%%%%%%%%%%%%%%%%%%%%%%%%%%%%%%%
\subsubsection{Patch model}

The patch density model describes the different density components
randomly distributed within the open field line area. The patch foots
located on the neutron star surface in the polar cap region can be
described as
\begin{eqnarray}
f(\theta)&=&f_0 \exp \left[-\frac{(\theta_\star - s
\theta_{\star{\rm lof}})^2}{2 \sigma_{\theta_\star}^2}\right],  \nonumber\\
g(\phi)&=&g_0 \exp \left[-\frac{(\phi-\phi_p)^2}{2
\sigma_{\phi}^2}\right].
\label{eq:patch_dis}
\end{eqnarray}
Here $g_0$ is the maximum density constant similar to $f_0$, $\phi_p$
is the peak position of a density patch in the magnetic azimuth
direction, and $\sigma_{\phi}$ is the corresponding characteristic
width. We do not use the subscript ``$_\star$'' for $\phi$ and $\phi_p$,
because they do not change with the height.

\begin{figure}
    \centering
    \includegraphics[angle=0, width=0.45\textwidth] {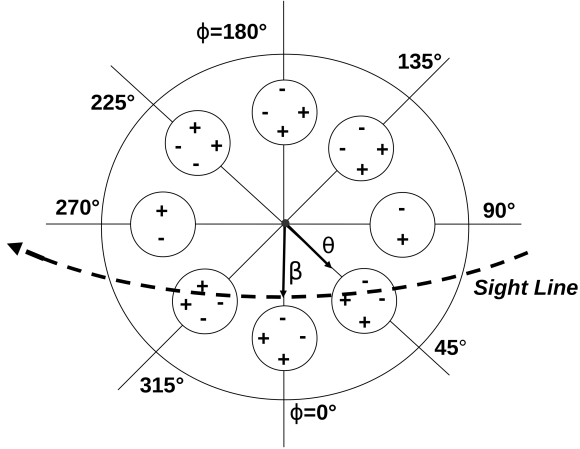}
    \caption{Sketch map of the discussed patches in the open field
      line region. The signs of "$+$" and "$-$" in the patches
      represent the signs of circular polarization at a height of 50
      $R_\star$. See detailed profiles in Fig.~\ref{fig:patch_phip0}
      and \ref{fig:patch_allphip}.}
\label{fig:patchsketch}
\end{figure}

The position of a patch ($s\theta_{\star\rm lof}$, $\phi_p$) affects the
density gradient along $\theta$ and $\phi$. For a given patch,
different cuts of a line of sight will cause different profiles. Here,
we calculate the profiles for typical patch positions in the open
field line region, with various impact
angles. Figure~\ref{fig:patchsketch} is the sketch map of the
positions of the discussed patches.

\begin{figure}
\centering
\includegraphics[angle=0, width=0.45\textwidth] {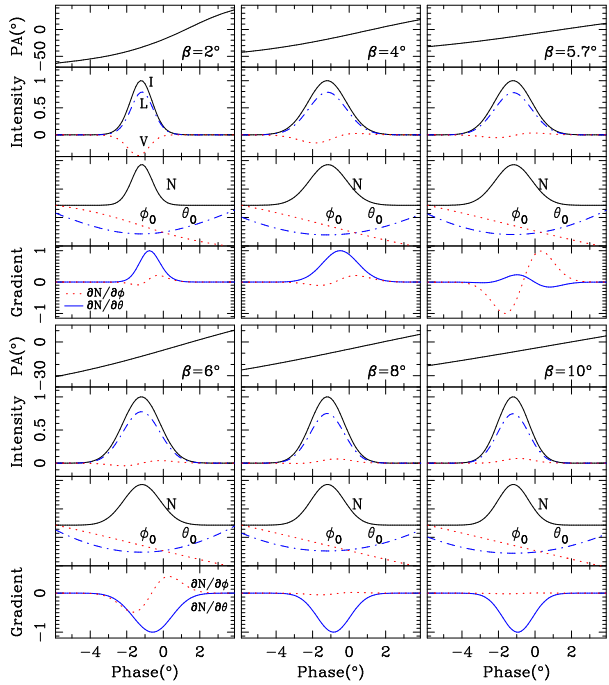}
\caption{Same as Fig.~\ref{fig:core} but for the density in a patch at
  $\phi_p=0$ with $r=50R_{\star}$. Profiles for six impact angles are
  shown. We plot $\partial N/\partial \phi/(\sin\theta)$ here instead
  of $\partial N/\partial \phi$, although marked as $\partial
  N/\partial \phi$ in the figure. The other parameters chosen here are
  $P=1s$, $\nu=400$\,MHz, $\alpha=30^{\circ}$, $\gamma=400$, $s=0.7$,
  $\sigma_{\theta_\star}=0.0006$ and $\sigma_\phi=0.1$.}
\label{fig:patch_phip0}
\end{figure}

\begin{figure}
\begin{tabular}{c}
\includegraphics[angle=0, width=0.42\textwidth] {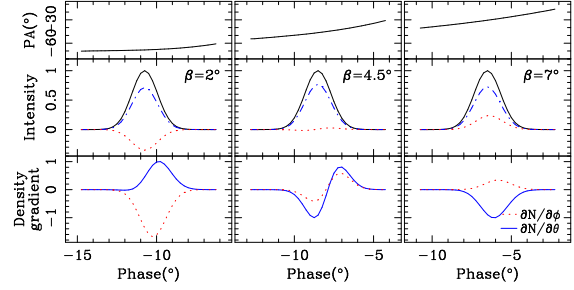} \\
                       (a) $\phi_p=45^{\circ}$               \\
                        \\
\includegraphics[angle=0, width=0.42\textwidth] {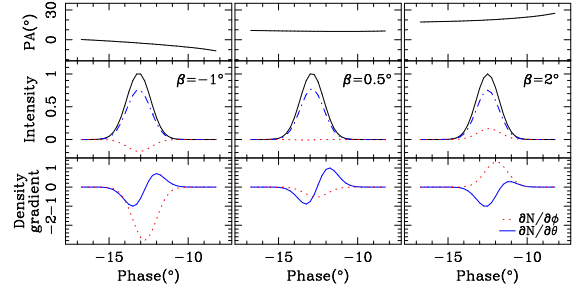}  \\
                       (b) $\phi_p=90^{\circ}$                \\
                         \\
\includegraphics[angle=0, width=0.42\textwidth] {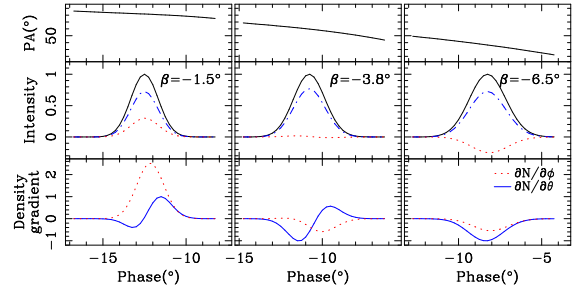} \\
                       (c) $\phi_p=135^{\circ}$               \\
                        \\
\includegraphics[angle=0, width=0.42\textwidth] {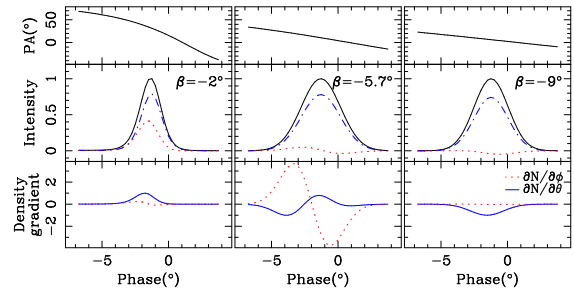} \\
                       (d) $\phi_p=180^{\circ}$               \\
                        \\
\includegraphics[angle=0, width=0.42\textwidth] {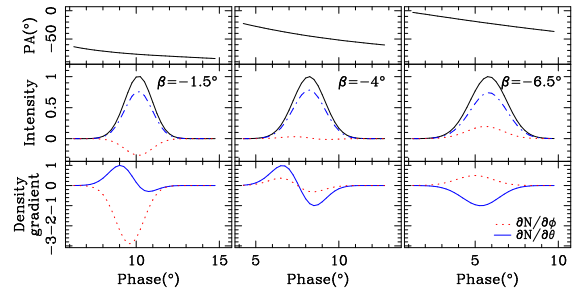} \\
                       (e) $\phi_p=225^{\circ}$               \\
\end{tabular}
\caption{Same as Fig.\ref{fig:patch_phip0} but for other patches in
  Fig.~\ref{fig:patchsketch}.}
\label{fig:patch_allphip}
\end{figure}

\addtocounter{figure}{-1}
\begin{figure}
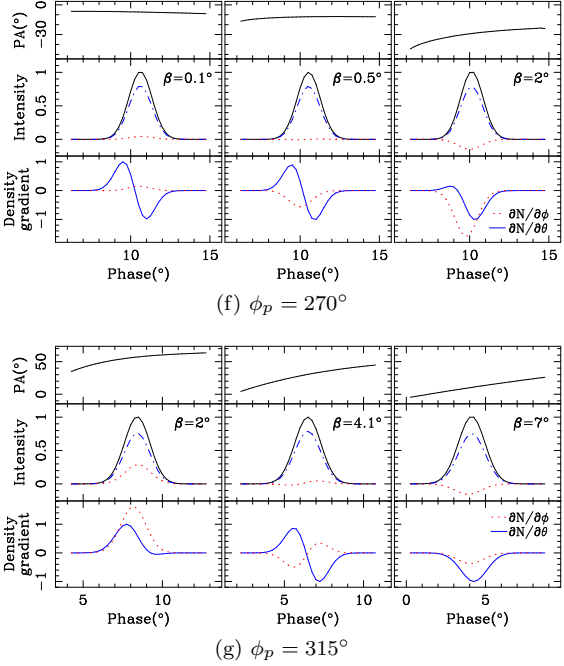

\begin{tabular}{c}
\includegraphics[angle=0, width=0.42\textwidth] {phi-90.ps} \\
                       (f) $\phi_p=270^{\circ}$               \\
                        \\
\includegraphics[angle=0, width=0.42\textwidth] {phi-45.ps}  \\
                       (g) $\phi_p=315^{\circ}$                \\
\end{tabular}
\caption{---continue}
\end{figure}

Firstly we consider a patch located in the meridional plane with
$\phi_p=0$. Sight lines cutting the patch with different $\beta$ give
different polarization profiles, as shown in
Fig.~\ref{fig:patch_phip0}. For $\beta=2^{\circ}$, the dominating
density gradient is $\partial N/\partial \theta$, which is large and
has a single sign at all the phases, and $\partial N/\partial\phi$ is
small. The circular polarization keeps to be of the right hand. When
$\beta$ increases to be $4^{\circ}$, there exits a little sign
reversal of CP in the trailing side of the profile, mainly because
$\partial N/\partial\phi$ is relatively strengthened. For the sight
line cuts across the patch center of $\beta=5.7^\circ$, $\partial
N/\partial \theta$ reaches its minimum, $\partial N/\partial \phi$
dominates. The weak circular polarization shows a sign reversal. For
large $\beta>5.7^\circ$, negative $\partial N/\partial \theta$
dominates again gradually, which gives the reversal of weak CP and
then the left hand CP.

Now, we consider the discrete patches (see Fig.~\ref{fig:patchsketch})
located at different positions around the magnetic axis with
$\phi_p=45^{\circ}$, $90^{\circ}$, $135^{\circ}$, $180^{\circ}$,
$225^{\circ}$, $270^{\circ}$ and $315^{\circ}$. The polarized profiles
from cutting through the inner, central and outer parts of the patch
are plotted for each patch in Fig.\ref{fig:patch_allphip}. Similar to
the patch of $\phi_p=0^\circ$, the $V$ profiles always have a single
sign, or the sign reversal, or the opposite sign for the line of sight
cutting the inner, central and outer parts of a patch. For the patches
at $\phi_p=90^\circ$ and $270^\circ$, we have $V\sim0$ when the line
of sight cuts across the central part of the patch. The $V$ profiles
for the patch at $\phi_p=180^\circ$ has almost the same dependence on
$\beta$ as the patch of $\phi_p=0^\circ$, except for the reversed
circular polarization. For other patches away from the meridional
plane, the $V$ profiles always depend on the variation of $\partial
N/\partial\phi$ and $\partial N/\partial\theta$ as listed in
Table~\ref{tab:Vsign}. But for patches with $\phi_p$ not close to
$0^\circ$ or $180^\circ$, the density gradient along $\phi$ dominates
the sign of circular polarization, because the rotation-induced
distortions are weak. For convenience, we mark the final sign of the
$V$ profile for different $\beta$ in each patch of
Fig.~\ref{fig:patchsketch}.

\begin{figure}
    \centering
    \includegraphics[angle=0, width=0.45\textwidth] {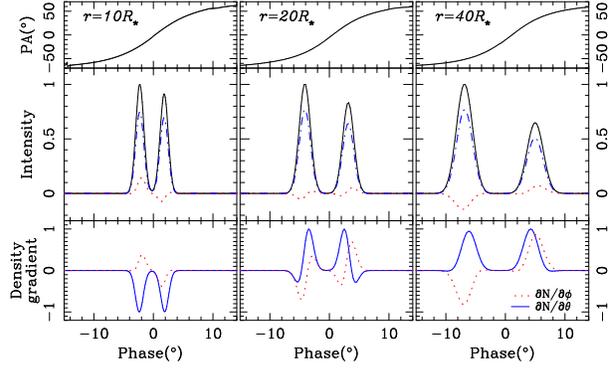}
    \caption{Same as Fig.\ref{fig:patch_phip0} but for two symmetrical
      patches ($\phi_p=\pm30^\circ$) at heights of 10$R_{\star}$,
      20$R_{\star}$ and 40$R_{\star}$. The impact angle is fixed to be
      $\beta=3^{\circ}$. The other parameters used here are $P=1s$,
      $\nu=400$\,MHz, $\alpha=30^{\circ}$, $\gamma=400$, $s=0.7$,
      $\sigma_{\theta_\star}=0.001$ and $\sigma_{\phi}=0.15$.}
\label{fig:patch_r}
\end{figure}

For a given patch, even when the impact angle $\beta$ is fixed, the
polarization profiles will be different for various emission
heights. The polarized emission from two symmetrical patches with a
fixed $\beta$ but different height (10$R_{\star}$, 20$R_{\star}$ and
40$R_{\star}$, respectively) are shown in
Fig.~\ref{fig:patch_r}. Naturally profiles gradually become wider with
the increasing of height. They also gradually shift towards an earlier
phase due to the aberration effect. The most important is the
evolution of the circular polarization with emission heights. At a low
height of 10$R_{\star}$, two components have different signs of CP,
and each with one sign only. At a height of 20$R_{\star}$ (the middle
panel), each component has the sign reversal of CP. At a height of
40$R_{\star}$, each component has one single sign of CP again without
reversal but the opposite hand compared to those at 10
$R_{\star}$. This is mainly because at different heights, the sight
lines with a fixed $\beta$ cut through the inner, middle and outer
parts of the patches, which should have different circular
polarization due to the dominating of $\partial N/\partial\phi$.

%%%%%%%%%%%%%%%%%%%%%%%%%%%%%%%%%%%%%%%%%%%%%%%%%%%%%%%%%%%%%%%%%%%%%%%%%%
%%%%%%%%%%%%%%%%%%%%%%%%%%%%%%%%%%%%%%%%%%%%%%%%%%%%%%%%%%%%%%%%%%%%%%%%%%
%%%%%%%%%%%%%%%%%%%%%%%%%%%%%%%%%%%%%%%%%%%%%%%%%%%%%%%%%%%%%%%%%%%%%%%%%%

\section{Emission from open field line region}

\begin{figure}
\centering
\includegraphics[angle=0, width=0.35\textwidth] {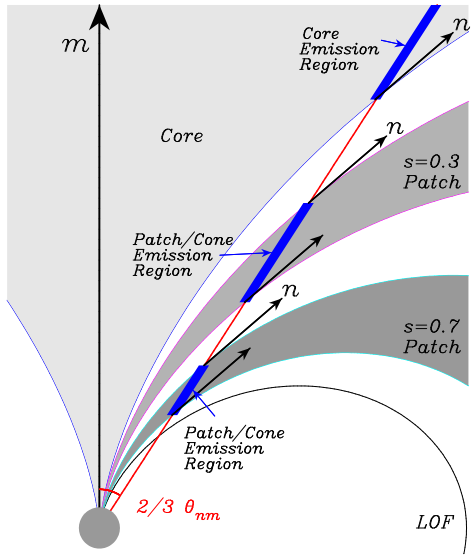}
\caption{Emission regions for different density models. The thick line
  outlines all the tangential emission points of field lines of
  $\theta\simeq2/3\theta_{\vecn\vecm}$ for a given line of sight, with
  $\theta_{\vecn\vecm}$ as the angle between wave vector $\vecn$ and
  magnetic momentum $\vecm$. The grey areas represent the emission
  regions for the core and cone/patch regions.}
\label{fig:emlinesketch}
\end{figure}

In general, a dipole field is assumed for pulsar magnetosphere. The
tangential points of the field lines of $\theta\simeq
2/3\,\theta_{\vecn\vecm}$ can produce emission detectable for a given
line of sight $\bmath n$. Here $\theta_{\vecn\vecm}$ is the angle of
the sight line from the magnetic axis $\vecm$, and
$\theta_{\vecn\vecm}=\beta$ when the sight line goes to the meridional
plane. Curvature radii are different for the tangential points at
different heights, so that characteristic emission frequencies are
different. This is the basic principle for the ``radius to frequency
mapping''. However, the emission from one height actually has a
wide-spread spectrum, the characteristic frequency is only the
frequency of the peak-intensity.

In fact, one receives the emission from the whole radiation region as
long as the line is of $\theta\simeq 2/3\,\theta_{\vecn\vecm}$ in the
open field line region (see Fig.~\ref{fig:emlinesketch}). The
retardation effect should be considered for the emission from
different heights. The emission profile from the whole open
field line region is given by
\begin{eqnarray}
I(\varphi)&=&\int I_r(\varphi+r/r_{\rm lc})\intd r, \nonumber\\
Q(\varphi)&=&\int Q_r(\varphi+r/r_{\rm lc})\intd r, \nonumber\\
U(\varphi)&=&\int U_r(\varphi+r/r_{\rm lc})\intd r, \nonumber\\
V(\varphi)&=&\int V_r(\varphi+r/r_{\rm lc})\intd r.
\label{eq:stokes_total}
\end{eqnarray}
Here $I_r$, $Q_r$, $U_r$ and $V_r$ are given by
equation~(\ref{eq:stokes_r}), and $r/r_{\rm lc}$ is the retarded phase
for the emission from different heights.

%%%%%%%%%%%%%%%    {patch emission}    %%%%%%%%%%%%%%%%%%%

\begin{figure}
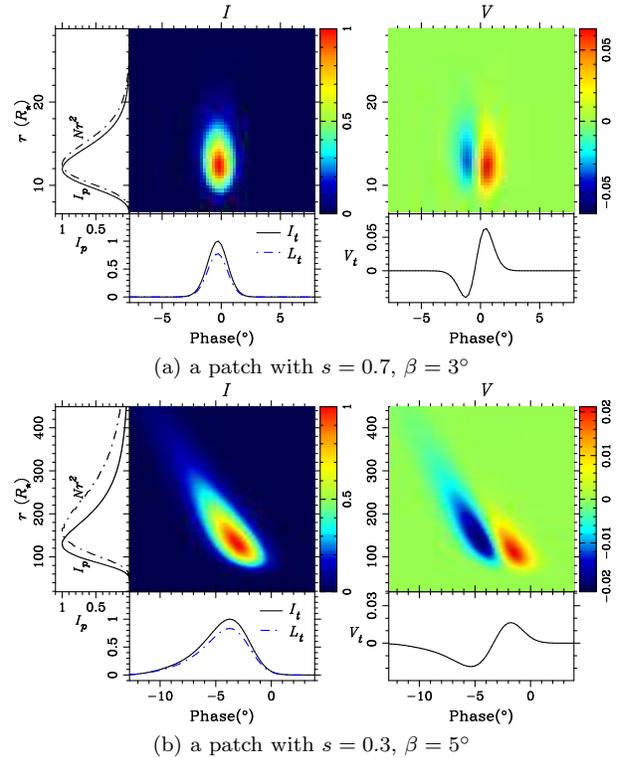

\centering
\begin{tabular}{c}
\includegraphics[angle=0, width=0.45\textwidth]
{patch_phase-height_distri.ps}          \\
(a) a patch with $s=0.7$, $\beta=3^\circ$  \\
\includegraphics[angle=0, width=0.45\textwidth]
{patch_phase-height_b5_s03_distri.ps}  \\
(b) a patch with $s=0.3$, $\beta=5^\circ$   \\
\end{tabular}
\caption{Total intensity and circular polarized emission of the
  curvature emission from two patches, plotted for various phases and
  heights. The linear polarization is very similar to the total
  intensity $I$ except for a slightly smaller magnitude. The
  integrated emission profiles are plotted under each map. The
  variations of the peak intensity $I_p$ (solid line) and $Nr^2$
  (dash-dotted line) are also plotted against heights in the left. The
  pulsar parameters used here are the same as those in
  Fig.~\ref{fig:patch_phip0}.}
\label{fig:It2D_patch}
\end{figure}

\begin{figure}
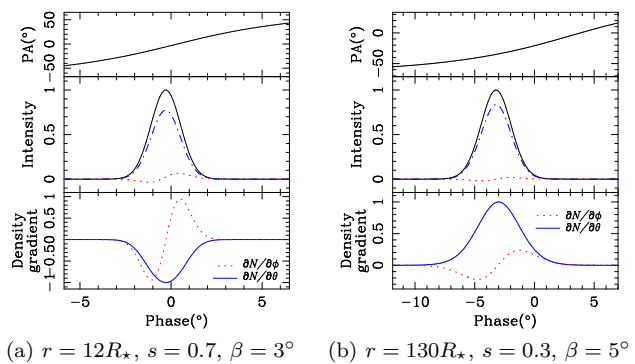

\centering
\begin{tabular}{cc}
\includegraphics[angle=0, width=0.21\textwidth] {patch_maxh_s07.ps}           &
\includegraphics[angle=0, width=0.21\textwidth] {patch_maxh_beta5_s03.ps} \\
(a) $r=12R_\star$, $s=0.7$, $\beta=3^\circ$                               &
(b) $r=130R_\star$, $s=0.3$, $\beta=5^\circ$                              \\
\end{tabular}
\caption{The emission profiles from the peak heights of the two
  patches in Fig.~\ref{fig:It2D_patch}. }
\label{fig:It_patch}
\end{figure}

The emission from a patch in the meridional plane with $s=0.7$ is
plotted in Fig.~\ref{fig:It2D_patch}a, with the polarization generated
at different phases and heights. The peak intensity $I_p$ (given by
  equation~\ref{eq:stokes_r}) varies with the heights. The significant
contribution to the integrated emission profiles in the bottom panel
comes from the region of 8--20$R_\star$. For $r<8R_\star$ and
$r>20R_\star$, the emission from this patch points away from the
observer (see the sketch map of Fig.~\ref{fig:emlinesketch}). The
dominating contribution comes from the peak height around
$r\simeq12R_\star$ for the maximum of plasma density (see the dashed
line in the left panel of Fig.~\ref{fig:It2D_patch}a). For the patch
with a higher and wider emission region of $s=0.3$
(Fig.~\ref{fig:It2D_patch}b), the dominating emission comes from the
height of about $130R_\star$. There the aberration and retardation
effects are rather strong, and the phase for the peak intensity and
CP are shifted to an early phase.

We plot the emission profiles from the peak height in
Fig.~\ref{fig:It_patch}, which are very similar to the integrated
total intensity and polarization profiles. The emission from the peak
height always has a sign reversal of the circular polarization,
because the density gradient along the magnetic azimuth $\phi$ has the
sign reversal. The aberration and retardation effects are very weak
for $r=12R_\star$ and the depolarization effect is also negligible.

\begin{figure}
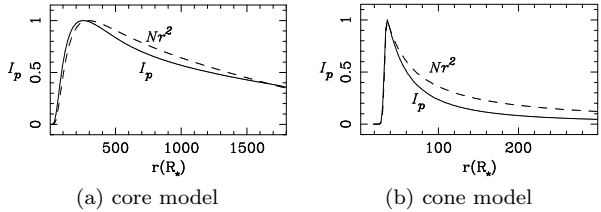

\centering
\begin{tabular}{cc}
\includegraphics[angle=0, width=0.21\textwidth]{core_Ipeak_gammaC.ps} &
\includegraphics[angle=0, width=0.21\textwidth]{cone_Ipeak_gammaC.ps} \\
(a) core model & (b) cone model \\
\end{tabular}
\caption{The peak intensity varies with the heights (solid lines) for
  core density model (panel a), and for cone model (panel b). The
  density variations, $N r^2$, are also plotted with the dashed
  lines. The parameters are the same as those in Fig.~\ref{fig:core}
  for core and Fig.~\ref{fig:cone} for cone.}
\label{fig:core-cone_Ipeak}
\end{figure}

We noticed that the polarization profiles from a patch are almost the
same for different frequencies. If the patch is wide and the density
gradient is small, e.g. the almost uniform distribution, the net
circular polarization from a patch should be very small. Note however
that the peak height is different for different frequencies, and the
integrated circular polarization may change for different frequencies.

%%%%%%%%%%%%%%%    {cone-core emission}    %%%%%%%%%%%%%%%%%%%

For the core density model, the emission region of
$\theta=2/3\theta_{\vecn\vecm}$ does not have an upper limit (see
Fig.~\ref{fig:emlinesketch}). The intensities from various emission
heights, up to thousands of $R_\star$, are still significant compared
with that from the peak height of $r\sim 250R_\star$ (see
Fig.~\ref{fig:core-cone_Ipeak}a). It is related to the integrand term
$Nr^2$ in equation~(\ref{eq:stokes_r}), which determines the final value of
$I_p$. The density variation can be written as
\begin{equation}
Nr^2=n(r)f(\theta)g(\phi)r^2\propto r^{-1}
\exp\left(-\frac{\theta^2}{2\sigma_{\theta_\star}^2}\frac{R_\star}{r}\right),
\end{equation}
and plotted by the dashed line in Fig.~\ref{fig:core-cone_Ipeak}a. The
total emission profile is very wide in phase because of the retarded
emission from large heights, and the polarization will also be
vanished.

\begin{figure}
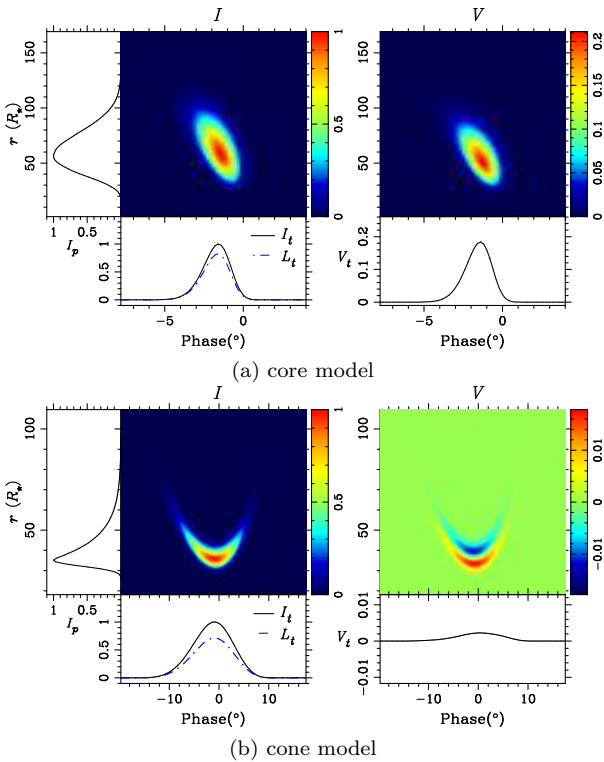

\centering
\begin{tabular}{c}
\includegraphics[angle=0, width=0.45\textwidth] {core_phase-height_distri.ps}\\
(a) core model          \\
\includegraphics[angle=0, width=0.45\textwidth] {cone_phase-height_distri.ps}\\
(b) cone model          \\
\end{tabular}
\caption{Same as Fig.~\ref{fig:It2D_patch}, except for the curvature
  emission from core (panel a) and cone (panel b). Lorentz factor
  damping along the particle trajectory is assumed as
  equation\ref{eq:lorentzr}, with an initial Lorentz factor at the star
  surface of $\gamma_0=1000$ and {\bf the damping factor
    $\kappa_{\gamma}=0.01$}.  The other parameters used for
  calculation here are the same as those in
  Fig.~\ref{fig:core-cone_Ipeak}.}
\label{fig:It2D_core-cone}
\end{figure}

In a practical case, the damping of the Lorentz factor of a particle
along $r$ is natural. For example, an exponential decay of the Lorentz
factor was introduced by \citet{qlz+01} to analyze the inverse Compton
scattering process in the magnetosphere. Here we take the slow
variation of the Lorentz factor with height as {\bf
\begin{equation}
\gamma=\gamma_{0}\exp(-\kappa_{\gamma} \frac{r-R_{\star}}{R_{\star}}),
\label{eq:lorentzr}
\end{equation}
}
where $\gamma_{0}$ is the energy for the secondary particles near the
neutron star polar cap, and {\bf $\kappa_{\gamma}$ is the damping factor}, and we
calculate the total emission profiles as shown in
Fig.~\ref{fig:It2D_core-cone}a. The emission region is constrained
to be between $20 R_{\star}$ and $120 R_{\star}$. The total polarized
profiles resemble to those from a height of $50 R_{\star}$
(Fig.\ref{fig:core}). The circular polarization contributed from each
height has the same sign, so the final total circular polarization has
only one single sign. The total intensity and polarization profiles
are asymmetry due to the retardation effect.

For the cone density model, the density $N r^2$ and the peak intensity
vary with height as shown in Fig.~\ref{fig:core-cone_Ipeak}b. The
emission can be produced even at a very large height. The dominating
emission comes from about 30 $R_{\star}$ to 300 $R_{\star}$. The
profiles become very wide (see Fig.\ref{fig:It2D_core-cone}b) because
of the strong retardation effect. With the damping of the Lorentz
factor of particles as the core model, the total emission profiles and
$I_p(r)$ are shown in Fig.~\ref{fig:It2D_core-cone}b. The cone density
only produces one profile component, instead of two. Because the
emission from low heights are very strong, and two components from a
higher regions are weak and only add the outer wings in both sides of
the profile from the lower height. The total circular polarization is
very weak also, because of the depolarization of the emission from
different heights.

%%%%%%%%%%%%%%%%%%%%%%%%%%%%%%%%%%%%%%%%%%%%%%%%%%%%%%%%%%%%%%%%%%%%%%
%%%%%%%%%%%%%%%%%%%%%%%%%%%%%%%%%%%%%%%%%%%%%%%%%%%%%%%%%%%%%%%%%%%%%%
%%%%%%%%%%%%%%%%%%%%%%%%%%%%%%%%%%%%%%%%%%%%%%%%%%%%%%%%%%%%%%%%%%%%%%
\section{Discussions and Conclusions}

In this paper, we depict the detailed scenery for curvature radiation
in rotating pulsar magnetosphere for the first time. The rotation
affects the velocity and acceleration of particles and hence the
emission property. Circular path approximation for the particle
trajectory is used at the emission point to calculate the radiating
electric field (see equation~\ref{eq:Ew}). The emission from particles at
the tangential points of field lines and the nearby field lines in the
$1/\gamma$ emission cone is considered for a given phase and
height. The polarization profiles from a given height and the whole
open field line region are calculated for three possible density
distributions in the form of core, cone and patches. We find the
following conclusions:
\begin{enumerate}
\item Rotation not only shifts the PA curves along the rotation phase,
  $\Delta\varphi_0$, but also causes an offset of the curve, $\PAi$,
  both of which are the first-order functions of the emission
  height. Its influences on $\alpha$ and $\beta$ determination follows
  the second order functions (Fig.~\ref{fig:pafit}).
\item Rotation distorts the patterns for the $1/\gamma$ cone more
  seriously if it originates from larger height and/or smaller impact
  angle (Fig.~\ref{fig:pattern}). The density gradients across the
  patterns will result in the net circular polarization.
\item For the patch density model, the $V$ profiles from the same
  height can have a single sign or the sign reversal or the opposite
  sign, depending on where the sight line is cutting
  (Fig.~\ref{fig:patch_phip0},\ref{fig:patch_allphip}). The
  polarization profiles resulting from a fixed impact angle also vary
  with emission heights (Fig.~\ref{fig:patch_r}).
\item The central peaked core will have the circular polarization of
  only one hand, which is usually significant and always has a single
  sign (Fig.~\ref{fig:It2D_core-cone}a). The cone emission at a large
  height is separated into two components with the opposite sign
  reversals for circular polarization. But at a lower height there is
  only one component with a single sign of $V$. The total circular
  polarization is very weak because of the depolarization of the
  emission from different heights (Fig.~\ref{fig:It2D_core-cone}b).
\end{enumerate}

Our calculations for the curvature radiation from particles
distributed in the form of core, cone and patches can be compared with
the observed pulsar polarization profiles. The sign reversal of
circular polarization are always observed in the central component of
the mean profile \citep{lm88,ran83,gs90}. \citet{hmx+98} pointed that
there also exists sign reversal of circular polarization for the cone
components for some pulsars. Such observational facts can be explained
if the emission comes from central or non-central patches. The CP with
a single sign may come from the core emission.

Here we did not consider the possible propagation effects, i.e., the
wave mode coupling effect \citep{wlh10} in the magnetosphere. Our
calculations can explain some pulsar polarized emission, but it can
not match various observation facts. It should be noted that we
consider just the curvature emission from particles with a single
$\gamma$ and assume the simple distributions of particle density. The
actual energy and density distributions of particles in the
magnetosphere are not clear at present. There exist other emission
mechanisms, like ICS \citep{xlh+00}, plasma process \citep{bgi88}.
Moreover, we treated the coherency in a simple way by assuming the
coherent bunch to be a huge point charge. However, the actual
coherent manner, bunch size and shape affect the radiation pattern
\citep{bb77}. We used the vacuum dipole magnetic field as the basic
model for our calculation, but there exists the possible distortion of
the dipole magnetic field due to rotation, which has not been taken
into account \citep{dh04}. The magnetic field structure, such as the
globally twisted, self-similar, force-free magnetosphere
\citep{tlk02}, can also lead to the modification of the radiation
pattern. All these will make the actual emission pattern much more
complicated than the simple calculations presented in this paper.

%%%%%%%%%%%%%%%%%%%%%%%%%%%%%%%%%%%%%%%%%%%%%%%%%%%%%%%%%%%%%%%%%%%%
\section*{Acknowledgments}

We thank R. T. Gangadhara and Dinesh Kumar for useful discussions and
suggestions. This work has been supported by the National Natural
Science Foundation of China (11003023 and 10833003).
%%%%%%%%%%%%%%%%%%%%%%%%%%%%%%%%%%%%%%%%%%%%%%%%%%%%%%%%%%%%%%%%
%%%%%%%%%%%%%%%%%%%%%%%%%%%%%%%%%%%%%%%%%%%%%%%%%%%%%%%%%%%%%%%%
%%%%%%%%%%%%%%%%%%%%%%%%%%%%%%%%%%%%%%%%%%%%%%%%%%%%%%%%%%%%%%%%
\bibliographystyle{mn2e}
\bibliography{ver}

\label{lastpage}

\end{document}